\documentclass[preprint,aps,epsfig]{revtex4}
\usepackage{graphicx,subfigure}
\usepackage{epsfig}
\usepackage{epstopdf}
\usepackage{cancel, color}

\def\lsim{\raise0.3ex\hbox{$<$\kern-0.75em\raise-1.1ex\hbox{$\sim$}}}
\def\gsim{\raise0.3ex\hbox{$>$\kern-0.75em\raise-1.1ex\hbox{$\sim$}}}

\def\pom{{I\!\!P}}
\newcommand{\be}{\begin{equation}}
\newcommand{\ee}{\end{equation}}

\def\beq{\begin{equation}}
\def\eeq{\end{equation}}
\def\beqa{\begin{eqnarray}}
\def\eeqa{\end{eqnarray}}

\newcommand{\rr}{\mbox{\boldmath $r$}}

\newcommand{\rb}{\mbox{\boldmath $b$}}

\def\gappeq{\mathrel{\rlap {\raise.5ex\hbox{$>$}}
{\lower.5ex\hbox{$\sim$}}}}

\def\lappeq{\mathrel{\rlap{\raise.5ex\hbox{$<$}}
{\lower.5ex\hbox{$\sim$}}}}

\def\Toprel#1\over#2{\mathrel{\mathop{#2}\limits^{#1}}}

\def\pom{{I\!\!P}}

\newcommand{\dd}{\, \mathrm{d}}
\begin{document}

\title{Nuclear shadowing in deep inelastic scattering on nuclei: a closer look} 
\author{ F. Carvalho$^1$, V.P.\ Gon\c{c}alves$^2$, F.S.\ Navarra$^3$ and  E.G.\ de~Oliveira $^3$}
\affiliation{$^1$Departamento de Ci\^encias Exatas e da Terra, Universidade Federal de S\~{a}o Paulo\\
Campus Diadema, Rua Prof. Artur Riedel, 275\\
Jd. Eldorado, 09972-270, Diadema, SP, Brazil\\
$^2$High and Medium Energy Group (GAME), \\
Instituto de F\'{\i}sica e Matem\'atica,  Universidade Federal de Pelotas\\
Caixa Postal 354, CEP 96010-900, Pelotas, RS, Brazil\\
$^3$Instituto de F\'{\i}sica, Universidade de S\~{a}o Paulo, 
C.P. 66318,  05315-970 S\~{a}o Paulo, SP, Brazil\\
}

\begin{abstract}

 The  measurement of the nuclear structure function $F_2^A (x,Q^2)$  at the future  electron-ion collider (EIC)  will be of great  
relevance to  understand the origin of the nuclear shadowing and to probe gluon saturation effects. 
 Currently there are several phenomenological models,  based on very distinct  approaches, which describe the scarce experimental data quite 
successfully. One of main uncertainties comes from the  schemes used to include the effects associated to the multiple 
scatterings and to unitarize the cross section. In this paper we compare the predictions of three distinct unitarization schemes of the 
nuclear structure function which use the same theoretical input to describe the projectile-nucleon interaction. In particular, 
we consider as input the predictions of the Color Glass Condensate formalism, which reproduce  the inclusive and diffractive $ep$ HERA data.
 Our results demonstrate that the experimental analysis of $F_2^A$ will be able to discriminate between the unitarization schemes.

\end{abstract}

\maketitle
\vspace{1cm}

\section{Introduction}

The measurement of the nuclear structure function in deep inelastic electron-nucleus scattering (DIS) is the best way to improve our knowledge of the nuclear parton distributions and QCD dynamics in the high energy regime (See, e.g.\cite{armesto_review,frank_review}). However, after more than 30 years of experimental and theoretical studies, a standard picture of nuclear modifications of structure functions and parton densities has not yet emerged. Fixed target DIS measurement on nuclei revealed that the ratio of nuclear to nucleon structure functions (normalized by the atomic mass number) is significantly different from unity.
In particular, these data demonstrate an intricate behavior, with the ratio being less than one at large $x$ (the EMC effect) and at small $x$ (shadowing) and larger than one for $x \approx 10^{-1}$ (antishadowing).  
  The existing data were taken at lower energies \cite{e665} and therefore  the perturbative QCD regime ($Q^2  \ge 1$ GeV$^2$) was explored  only for relatively large values of the (Bjorken) $x$  variable ($x > 10^{-2} $). Experimentally, this situation will hopefully change with  a future high energy electron-ion collider (EIC) (For recent reviews see, e.g. \cite{erhic,lhec}),  which is supposed to take data at higher energies and explore the region of small $x$ ($   x < 10^{-2} $) in the perturbative QCD regime.  
  
The theory of nuclear effects in DIS is still far from being concluded. The straightforward use of nucleon parton distributions evolved with 
DGLAP equations and corrected with a nuclear  modification factor  determined by fitting the existing data as in 
Refs. \cite{eks1,eks2,hkn,ds,eps1,eps2} 
is  well justified only in the large $Q^2$ region and not too small $x$. Moreover, these approaches do not address the fundamental problem of 
the origin of the nuclear shadowing and  cannot be extended to small $x$,  where we expect to see  new interesting physics related to the 
non-linear aspects of QCD and gluon saturation (For  reviews see Ref. \cite{hdqcd}). Currently, there are several phenomenological models which 
predict different magnitudes for the shadowing in the nuclear structure function  based on distinct  treatments for  the multiple scatterings of the partonic 
component of the virtual photon, assumed in general to be a  quark-antiquark ($q \bar{q}$) color dipole. 
Some works \cite{armesto_glauber,erike_inclusive,simone_hq,erike_exclusive} address the origin of the nuclear shadowing through the Glauber-Gribov formalism \cite{glauber,gribov} in the totally coherent limit ($l_c \approx 1 / 2 m_N x \gg R_A$, where $l_c$ is coherence length), which considers the multiple scattering of the color dipole with a nucleus made  of nucleons whose binding energy is neglected. In the high energy limit, the eikonal approximation is assumed, with the dipole keeping a fixed size during the scattering process.  In this approach the total photon-nucleus cross section is given by 
\beq
\sigma_{\gamma^*A} = \int d^2r \, \int dz |\psi(r,z)|^2 \sigma_{dA}(x,r)
\label{sigga}
\eeq
where $|\psi(r,z)|^2$ is the probability of the photon to split into a $q\bar{q}$ pair of size $r$ and   
$\sigma_{d A}(x,r)$ is the dipole-nucleus cross section, which is expressed as \cite{armesto_glauber}
\beq
\sigma_{dA}(x,r) = \int d^2b \,2 \, \left[ 1-\exp\left(-\frac{1}{2}A \, T_A(b) \sigma_{dp}(x,r)\right) \right]
\label{sigda}
\eeq
with $T_A(b)$ being  the nuclear thickness function and  $\sigma_{dp}(x,r)$ is the dipole-proton cross section.
It must be stressed that once $\sigma_{dp}(x,r)$ is fixed, the extension to the nuclear case is essentially parameter free in this approach. 
In the Glauber formula (\ref{sigda}) it is assumed  that the dipole undergoes several elastic scatterings on the target. Although reasonable 
and phenomenologically successful this assumption deserves further investigation.  
This model can be derived in the classical approach of the Color Glass Condensate formalism \cite{raju_acta}.

Another approach largely used in the literature is 
 based on the  connection between nuclear shadowing and the cross section for the diffractive dissociation 
of the projectile \cite{capella,frank_nuc,armesto_alba,armesto_dif}, which was established  long time ago 
by Gribov \cite{gribov}. Its result can be derived using reggeon calculus \cite{reggeon} and the Abramovsky-Gribov-Kancheli (AGK) cutting rules \cite{agk} and is a manifestation of the unitarity. This formalism can be used to calculate directly cross sections of photon-nucleus scattering for the interaction with two nucleons in terms of the diffractive photon-nucleon cross section.
In this formalism, the total photon-nucleus cross section is expressed as  a series 
 containing the contribution from multiple scatterings (1, 2, $\dots$):
\beq
\sigma_{\gamma^*A} = \sigma_{\gamma^*A}^{(1)} + \sigma_{\gamma^*A}^{(2)} + \sigma_{\gamma^*A}^{(3)} + \cdots\,
\label{eq1}
\eeq
with the first term being the one that arises from independent scattering of the photon off $A$ nucleons: 
\beq
\sigma_{\gamma^* A}^{(1)}=A\,\sigma_{\gamma^*p}
\eeq
and the first correction to the non-additivity of cross sections being
\beq
\sigma_{\gamma^* A}^{(2)}=-4\pi A(A-1)\int d^2b\ T_A^2(b)
\int _{M^2_\mathrm{min}}^{M^2_\mathrm{max}}dM^2 \left.
\frac{d\sigma^{\mathcal{D}}_{\gamma^*{\rm p}}}{dM^2dt}\right\vert_{t=0} F_A^2(t_\mathrm{min})
\label{sigalb} 
\eeq
where $M^2$ is the mass of the diffractively produced system, $F_A$ is the nucleus form factor which takes into account the coherence effects and the differential $\gamma^* p$ cross section for diffractive dissociation of the virtual photon  appearing in 
(\ref{sigalb}) is given by: 
\beq
\left.\frac{d\sigma^\mathcal{D}_{\gamma^*{\rm p}} (Q^2,x_{\pom},\beta)}{dM^2dt}
\right\vert_{t=0}=
\frac{4\pi^2\alpha_{em}B_D}{Q^2(Q^2+M^2)}x_{\pom}F^{(3)}_{2\mathcal{D}}(Q^2,x_{\pom},\beta)
\label{eq3}
\eeq
where $B_D$ is the diffractive slope parameter and $x_{\pom}F^{(3)}_{2\mathcal{D}}(Q^2,x_{\pom},\beta)$ is the diffractive proton structure function. Moreover,  
$t_\mathrm{min}=-m_N^2 x_\mathcal{P}^2$, $x_{\pom}=x/\beta$ and $\beta=Q^2/(Q^2+M^2)$. 
The integration limits in $M^2$ are  $M^2_\mathrm{min}= 4 m_\pi^2 =0.08$ GeV$^2$, $M^2_\mathrm{max}= Q^2\left(x_{\pom \mathrm{max}}/x-1\right)$ and $x_{{\pom}\mathrm{max}}=0.1$. 
A shortcoming of this approach is that the inclusion of the higher order rescatterings is model dependent.  This resummation is specially important at small $x$, where multiple scattering is more likely to happen.
 In general it is assumed that the intermediate states in the rescatterings have the same structure and two resummation schemes are considered: (a) {\it the Schwimmer equation} \cite{schwimmer}, which sums all fan diagrams with triple pomeron interactions and which is valid for the scattering of a small projectile on a large target. It implies that  the photon--nucleus cross section is given by:
\beq
\sigma_{\gamma^* A}^{S}(x,r)=\sigma_{\gamma^*p}(x,r) \, A \,  \int d^2b \frac{   T_A(b)}{1+(A-1) \,  T_A(b) \, f(x,Q^2)}
\label{schwimmer}
\eeq
and (b) {\it the eikonal unitarized cross section}, given by
\beq
\sigma^{E}_{\gamma^*A}(x,r)=\sigma_{\gamma^*p} (x,r) \, A \,  \int d^2b 
 \frac{\left\{1-\exp{\left[-2(A-1)T_A(b)f(x,Q^2)\right]}\right\}}{2(A-1)f(x,Q^2)},
\label{eikonal}
\eeq
where
\beq
f(x,Q^2)=\frac{4\pi}{\sigma_{\gamma^*p}(x,r)} \times \int_{M^2_{min}}^{M^2_\mathrm{max}} dM^2 \left. \frac{d\sigma^\mathcal{D}}{dM^2 dt} \right|_{t=0} \times F^2_A(t_\mathrm{min}).
\label{ff}
\eeq
As shown in \cite{armesto_alba,armesto_dif},  the eikonal unitarization predicts a larger magnitude for the nuclear shadowing than the Schwimmer equation. 
For models which take into account the possibility of different intermediate states see, e.g., Ref. \cite{kope}. 
Except for the choice of the resummation scheme, the predictions for $\sigma_{\gamma^* A}$ obtained using (\ref{schwimmer}) or (\ref{eikonal}) are parameter free once the diffractive cross section is provided. Models based on this non-perturbative Regge-Gribov framework are quite successful in describing existing data on inclusive and diffractive $ep$ and $eA$ scattering \cite{armesto_dif,armesto_ep}. However, they lack solid theoretical foundations within QCD. 
It is important to emphasize that some authors \cite{frank_nuc} use these models as initial conditions for DGLAP evolution.

The comparison among the predictions of the different models for nuclear shadowing presented in Ref. \cite{armesto_review},  including the models discussed above, shows that they coincide within $\approx 15 \%$ in the region where experimental data exist ($x \ge 10^{-2}$) but  differ strongly for smaller values of $x$, with the difference being almost of a factor 2 at $x = 10^{-5}$. Our goal in this paper is try to reduce the theoretical uncertainty present in these predictions. In particular, 
differently from previous studies, which consider different inputs in the calculations using the Glauber, Schwimmer and Eikonal approaches, we will consider a unique model for the projectile - nucleon interaction. We will calculate the dipole - nucleon cross section and the diffractive structure function  using the dipole picture and the solution of the running coupling Balitsky-Kovchegov equation \cite{bk}, which is the basic equation of the Color Glass Condensate formalism. Recently, this approach 
was shown to describe quite well the $ep$ HERA data for inclusive and diffractive observables (See, e.g. Refs. \cite{rcbk,vic_joao,alba_marquet,vic_anelise}). 
Following this procedure we are able to estimate the magnitude of the theoretical uncertainty associated to the way the multiple scatterings are considered, reducing the contribution associated to the choice of initial conditions used in the calculations. Moreover, we discuss the possibility of discriminating   
between these unitarization procedures in a future electron-ion collider.

This paper is organized as follows. In Sec. \ref{dipole} we present a brief description of inclusive and diffrative $\gamma$ - nucleon processes in the color dipole picture with particular emphasis in the dipole - proton cross section given by the Color Glass Condensate formalism. In Section \ref{results} we present the predictions of the three unitarization schemes discussed above using as input  the CGC results for the dipole - proton interaction and compare them 
with the existing experimental data. Moreover, we present a comparison between the predictions for the kinematical region which will be probed in a future electron - ion collider. Finally, in Section \ref{conc} we summarize our results and present our conclusions.

\section{Inclusive and diffractive $\gamma p$ processes in the color dipole picture}
\label{dipole}

The photon-hadron interaction at high energy (small $x$) is usually described in the infinite momentum frame  of the hadron in terms of the scattering of the photon off a sea quark, which is typically emitted  by the small-$x$ gluons in the proton. However, as already mentioned in the introduction, in order to 
describe inclusive and diffractive interactions and disentangle the small-$x$ dynamics of the hadron wavefunction, it is more adequate to consider the photon-hadron scattering in the dipole frame, in which most of the energy is
carried by the hadron, while the  photon  has
just enough energy to dissociate into a quark-antiquark pair
before the scattering. In this representation the probing
projectile fluctuates into a
quark-antiquark pair (a dipole) with transverse separation
$\rr$ long before the interaction, which then
scatters off the target \cite{dipole}. The main motivation to use this color dipole approach is that it gives a simple unified picture of inclusive and diffractive processes. In particular,
in this approach the proton structure function is given in terms of the dipole - proton cross section, $\sigma_{d p}(x,r)$, as follows:
\beq
F_2^p(x,Q^2) = \frac{Q^2}{4\pi^2\alpha_{em}}   \int d^2r \, \int dz |\psi(r,z)|^2 \sigma_{dp}(x,r)
\label{sigga}
\eeq
where $|\psi(r,z)|^2$ is the probability of the photon to split into a $q\bar{q}$ pair of size $r$. Moreover, the total diffractive cross sections take the following form  (See e.g. Ref. \cite{GBW}),
\begin{equation}\label{eq:sigdiff}
\sigma^\mathcal{D}_{T,L} = \int_{-\infty}^0 dt\,e^{B_D t} \left. \frac{d \sigma ^\mathcal{D} _{T,L}}{d t} \right|_{t = 0} = \frac{1}{B_D} \left. \frac{d \sigma ^\mathcal{D} _{T,L}}{d t} \right|_{t = 0}
%\label{sig_difra}
\end{equation}
where 
\begin{equation}\label{eq:dsig-dt}
\left. \frac{d \sigma ^\mathcal{D} _{T,L}}{d t} \right|_{t = 0} = \frac{1}{16 \pi} \int d^2 {\bf r} 
\int ^1 _0 d \alpha |\Psi _{T,L} (\alpha, {\bf r})|^2 \sigma _{dp} ^2 (x, \rr) 
\end{equation}
It is assumed that the dependence on the momentum transfer, $t$, factorizes and is given by an exponential with diffractive slope $B_D$. 
The diffractive processes can be analysed in more detail by studying the behaviour of the diffractive structure function $F_2^{\mathcal{D} (3)}(Q^{2}, \beta, x_{I\!\!P})$. Following Ref. \cite{GBW} we assume that the diffractive structure function is given by
\begin{equation}
F_2^{\mathcal{D}(3)} (Q^{2}, \beta, x_{I\!\!P}) = F^{\mathcal{D}}_{q\bar{q},L} + F^{\mathcal{D}}_{q\bar{q},T} + F^{\mathcal{D}}_{q\bar{q}g,T},
\label{soma}
\end{equation}
where the $q\bar q g$ contribution with longitudinal polarization is not
present because it has no leading logarithm in $Q^2$. The different contributions can be calculated and for the $q\bar q$ contributions
they read \cite{wusthoff,nikqqg}
\begin{equation}
  x_{I\!\!P}F^{\mathcal{D}}_{q\bar{q},L}(Q^{2}, \beta, x_{I\!\!P})=
\frac{3 Q^{6}}{32 \pi^{4} \beta B_D} \sum_{f} e_{f}^{2} 
 2\int_{\alpha_{0}}^{1/2} d\alpha \alpha^{3}(1-\alpha)^{3} \Phi_{0},
\label{qqbl}
\end{equation}
\begin{equation}
 x_{I\!\!P}F^{\mathcal{D}}_{q\bar{q},T}(Q^{2}, \beta, x_{I\!\!P}) =  
 \frac{3 Q^{4}}{128\pi^{4} \beta B_D}  \sum_{f} e_{f}^{2} 
 2\int_{\alpha_{0}}^{1/2} d\alpha \alpha(1-\alpha) 
\left\{ \epsilon^{2}[\alpha^{2} + (1-\alpha)^{2}] \Phi_{1} + m_f^{2} \Phi_{0}  \right\}   
\label{qqbt}
\end{equation}
where the lower limit of the integral over $\alpha$ is given by $\alpha_{0} = \frac{1}{2} \, \left(1 - \sqrt{1 - \frac{4m_{f}^{2}}{M^{2}}}\right)
$, the sum is performed over the quark flavors and \cite{fss}
\begin{equation}
\Phi_{0,1}  \equiv  \left(\int_{0}^{\infty}r dr K_{0 ,1}(\epsilon r)\sigma_{dp}(x_{I\!\!P},\rr) J_{0 ,1}(kr) \right)^2.
\label{fi}
\end{equation}
The $q\bar{q}g$ contribution, within the dipole picture at leading $\ln Q^2$ accuracy, is given by \cite{wusthoff,GBW,nikqqg}
 \begin{eqnarray}
   \lefteqn{x_{I\!\!P}F^{\mathcal{D}}_{q\bar{q}g,T}(Q^{2}, \beta, x_{I\!\!P}) 
  =  \frac{81 \beta \alpha_{S} }{512 \pi^{5} B_D} \sum_{f} e_{f}^{2} 
 \int_{\beta}^{1}\frac{\mbox{d}z}{(1 - z)^{3}} 
 \left[ \left(1- \frac{\beta}{z}\right)^{2} +  \left(\frac{\beta}{z}\right)^{2} \right] } \label{qqg} \\
  & \times & \int_{0}^{(1-z)Q^{2}}\mbox{d} k_{t}^{2} \ln \left(\frac{(1-z)Q^{2}}{k_{t}^{2}}\right) 
\left[ \int_{0}^{\infty} u \mbox{d}u \; \sigma_{dp}(u / k_{t}, x_{I\!\!P}) 
   K_{2}\left( \sqrt{\frac{z}{1-z} u^{2}}\right)  J_{2}(u) \right]^{2}.\nonumber
\end{eqnarray} 
As pointed in Ref. \cite{marquet}, at small $\beta$ and low $Q^2$, the leading $\ln (1/\beta)$ terms should be resummed and the above expression should be modified. However, as a description with the same quality using the Eq. (\ref{qqg}) is possible by adjusting the coupling \cite{marquet}, in what follows we will use this expression for our phenomenological studies. 
We  use the standard notation for the variables  $
x_{I\!\!P} = (M^2 + Q^2)/(W^2 + Q^2)$ and $x = Q^2/(W^2 + Q^2) = \beta x_{\pom}$, 
where  $W$ the total energy of the 
$\gamma ^* p$ system.

The main input for the calculations of  inclusive and diffractive observables in the dipole picture is $\sigma_{dp}(x,\rr)$ which is determined by the QCD dynamics at small $x$. In the eikonal approximation, it is  given by:
\begin{equation} 
\sigma_{dp}(x, \rr) = 2 \int d^2 \rb \,  {\cal N}(x, \rr, \rb)
\label{sdip}
\end{equation}
where $ {\cal N}(x, \rr, \rb)$ is the forward scattering amplitude for a dipole with size 
$r=|\rr|$ and impact parameter $\rb$  which can be related to the expectation value of a Wilson loop \cite{hdqcd}. It
encodes all the
information about the hadronic scattering, and thus about the
non-linear and quantum effects in the hadron wave function. In general, it is  assumed that the impact parameter dependence of $\cal{N}$ can be factorized as  ${\cal{N}}(x,\rr,\rb) = {\cal{N}}(x,\rr) S(\rb)$, where
$S(\rb)$ is the profile function in impact parameter space, which implies  $\sigma_{dp}(x,\rr)=\sigma_0 \mathcal{N}(x,\rr)$. The forward scattering amplitude ${\cal{N}}(x,\rr)$ 
can be obtained by solving the BK evolution equation \cite{rcbk} or considering phenomenological QCD inspired models to describe the interaction of the dipole with the target. BK equation is the simplest nonlinear evolution equation
for the dipole-hadron scattering amplitude, being actually a mean field version
of the first equation of the B-JIMWLK hierarchy \cite{CGC}. In its linear
version, it corresponds to the Balitsky-Fadin-Kuraev-Lipatov (BFKL) equation
\cite{bfkl}.
The solution of the LO BK equation implies that the saturation scale grows much faster with increasing energy
($Q_s^2\sim x^{-\lambda}$, with $\lambda \approx 0.5$) than that
extracted from phenomenology ($\lambda \sim 0.2-0.3$). 

In the last years the next-to-leading order corrections to the  BK equation were
 calculated  
\cite{kovwei1,javier_kov,balnlo} through the ressumation of $\alpha_s N_f$ contributions to 
all orders, where $N_f$ is the number of flavors. Thanks to these works it is now possible  to estimate 
the soft gluon emission and running coupling corrections to the evolution kernel.
The authors have found out  that  the dominant contributions come from the running 
coupling corrections, which allow us to  determine the scale of the running coupling in the 
kernel. The solution of the improved BK equation was studied in detail in Ref. 
\cite{javier_kov}. The running of the coupling reduces 
the speed of the evolution to values compatible with experimental data, with the geometric 
scaling regime being reached only at ultra-high energies. In \cite{rcbk} a global 
analysis of the small $x$ data for the proton structure function using the improved BK 
equation was performed  (See also Ref. \cite{weigert}). In contrast to the  BK  equation 
at leading logarithmic $\alpha_s \ln (1/x)$ approximation, which  fails to describe the 
HERA data, the inclusion of running coupling effects in the evolution renders the BK equation 
compatible with them (See also \cite{vic_joao,alba_marquet,vic_anelise}). In what follows we 
consider the BK predictions for ${\cal{N}}(x,\rr)$ (from now on called rcBK) obtained using the GBW \cite{rcbk} initial 
condition.

%\begin{figure}
%\begin{tabular}{cc}
%\includegraphics[scale=0.15]{PbD-rcBK2.eps} & \includegraphics[scale=0.15]{CaD-rcBK2.eps}
%%\psfig{file=hqpp_charm_rhic.eps,width=80mm} & \psfig{file=hqpp_bottom_rhic.eps,width=80mm}
%\end{tabular}
%\caption{Comparison between the predictions of the distinct models and the E665 experimental data at small $x$. }
%\label{fig1}
%\end{figure}

\begin{figure}
\includegraphics[scale=0.20]{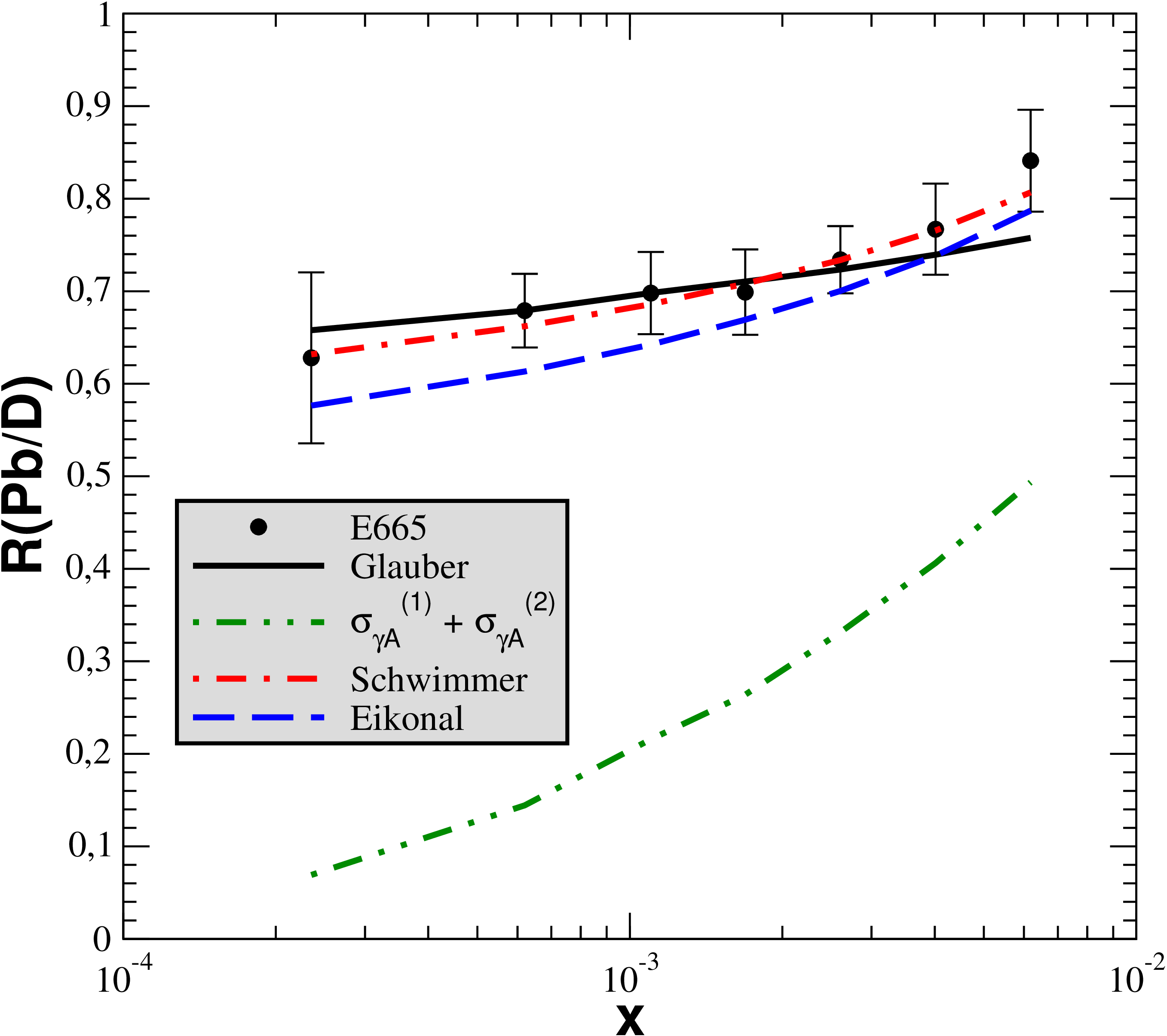} 
\caption{Comparison between the predictions of the distinct models and the E665 experimental data at small $x$. }
\label{fig1}
\end{figure}

%--------------------------------------------------------------------------------
\section{Numerical results and discussion}
\label{results}
In what follows we shall  consider two different nuclei, Ca and Pb, and use the deuteron (D) as a reference to calculate the experimentally measured ratios $R_{Ca/D} \equiv (2/40) F_2^{Ca}/F^D_2$ and $R_{Pb/D} \equiv (2/208)F^{Pb}_2 / F^D_2$.
We assume that the diffractive slope parameter is $B_D = 6.7$ GeV$^{-2}$ and that  the nucleus form factor is given by:
\beq
F_A(t_\mathrm{min})=\int d^2b\ J_0(b\sqrt{-t_\mathrm{min}})T_A(b),
\label{eq2-1}
\eeq
where the thickness function is given in terms of the nuclear density $\rho_A$ as:
$$
T_A(b)=\int_{-\infty}^{+\infty} dz\rho_A(\vec{b},z),
$$
with the normalization fixed by $ \int d^2b\ T_A(\vec{b})=1$. 

In Fig. \ref{fig1} we compare the predictions   of the Glauber (solid line), Schwimmer (dot-dashed line), Eikonal (dashed line) and double scattering (dot-dot-dashed line)  models for the ratios with the E665 experimental data at small $x$ \cite{e665}. 
 Although  joined with lines, our results are computed at the same $\langle x \rangle$ and $\langle Q^2 \rangle$ as the experimental data.  Our results demonstrate that if we compute the nuclear structure function up to two scatterings, which implies that 
$\sigma_{\gamma^*A} = \sigma_{\gamma^*A}^{(1)} + \sigma_{\gamma^*A}^{(2)}$, we are not able to describe the experimental data. Furthermore, since the 
magnitude of the first correction, $\sigma_{\gamma^*A}^{(2)}$, is very large, 
then there is no hope to estimate the nuclear structure function by just summing a few terms in the multiple scattering series. Therefore, a full resummation of the multiple scatterings is necessary, which makes the predictions model dependent.
The agreement with the current  experimental data at small $x$ of the Glauber, Schwimmer and Eikonal models is quite reasonable taking into account that no parameters have been fitted to reproduce the data. This implies that the current data are not able to discriminate between the unitarization schemes.

\begin{figure}
\begin{tabular}{cc}
\includegraphics[scale=0.15]{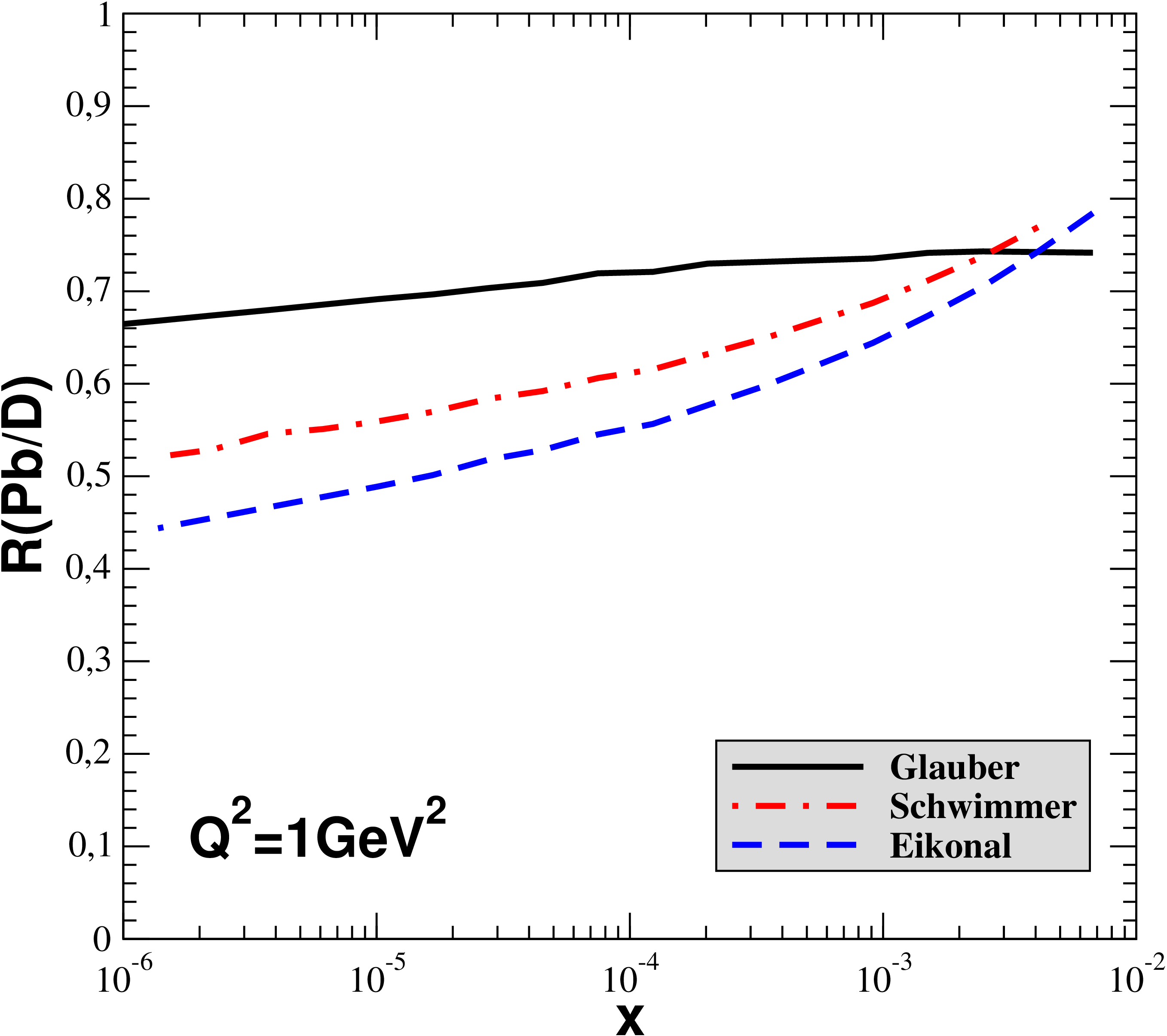} & \includegraphics[scale=0.15]{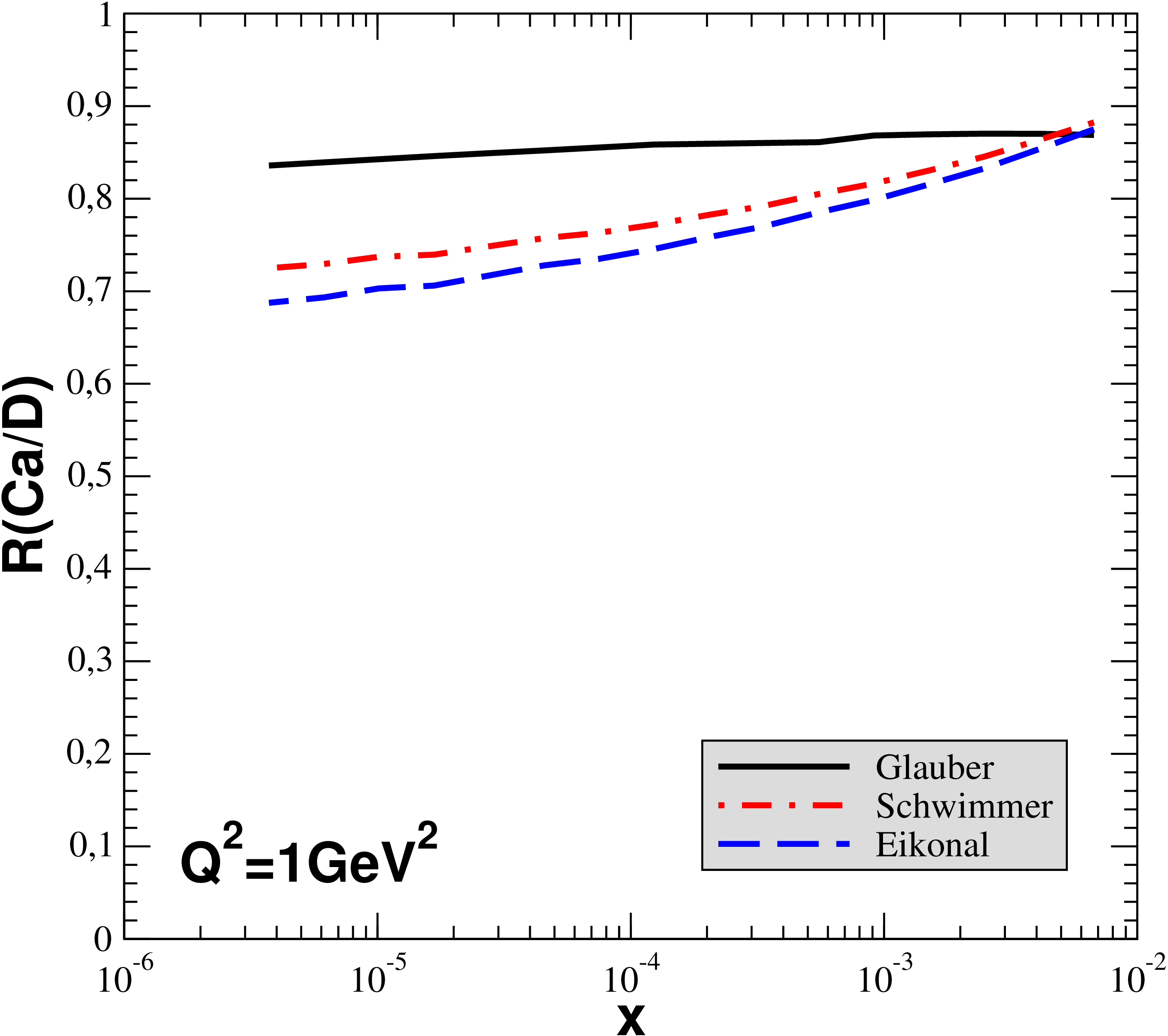}
\end{tabular}
\caption{Nuclear ratios $R_{Pb/D}$ (left panel) and $R_{Ca/D}$ (right panel) as a function of $x$ at $Q^2 = 1$ GeV$^2$.}
\label{fig2}
\end{figure}

% ratioPbDwitheps09.eps

Having in mind that a future electron - ion collider is expected to be able to analyse the kinematical region of 
small $x$ ($x \simeq 10^{-5}$) and $Q^2 \ge 1$ GeV$^2$,  we now compute the ratios $R_{Ca/D}$ and $R_{Pb/D}$ as a function of $x$ for  two different values of $Q^2$ (= 1 and 10 GeV$^2$). In Fig. \ref{fig2} we present our predictions for $Q^2 = 1$ GeV$^2$.
It is important to emphasize that in electron scattering the range of $x$-values attainable is kinematically restricted to $x > Q^2/s$, where $s$ is the squared center-of-mass energy, which implies that at $Q^2 = 1$ GeV$^2$ the smaller values of $x$ in the perturbative region will be probed.  At large $x$ ($\approx 10^{-2}$) the predictions almost coincide. However, at small $x$,  the predictions  based on the Schwimmer equation or on  the eikonal unitarized cross section 
give a stronger shadowing than those based on Glauber-like rescatterings.  In particular, at $x \approx 10^{-4}$, the difference between Glauber and Schwimmer is almost  10 \% in the ratio  $R(Ca/D)$ increasing to $\approx$ 20 \% in $R(Pb/D)$. At this $x$ value, the difference between Schwimmer and Eikonal is $\approx$ 5 \% and 12 \% for the ratios $R(Ca/D)$ and $R(Pb/D)$, respectively.
At smaller values of $x$, the difference between the three  predictions increases, being  larger than 20\%.
 Consequently, a measurement of $F_2^A$ at $A = Pb$ at small $x$ with $\approx 10 \%$ precision would be a sensitive test to discriminate between the 
different models.
 
\begin{figure}
\begin{tabular}{cc}
\includegraphics[scale=0.15]{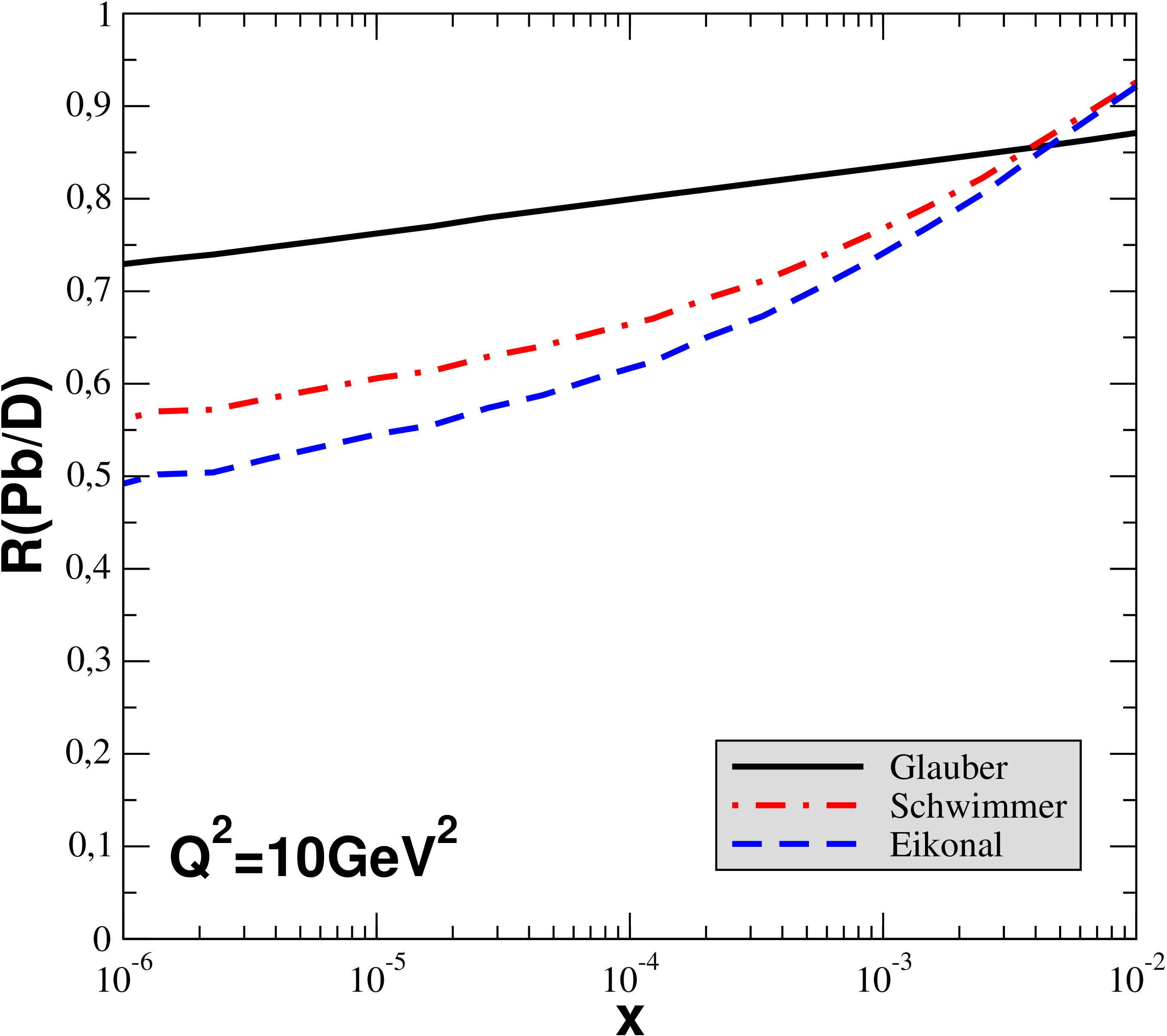} & \includegraphics[scale=0.15]{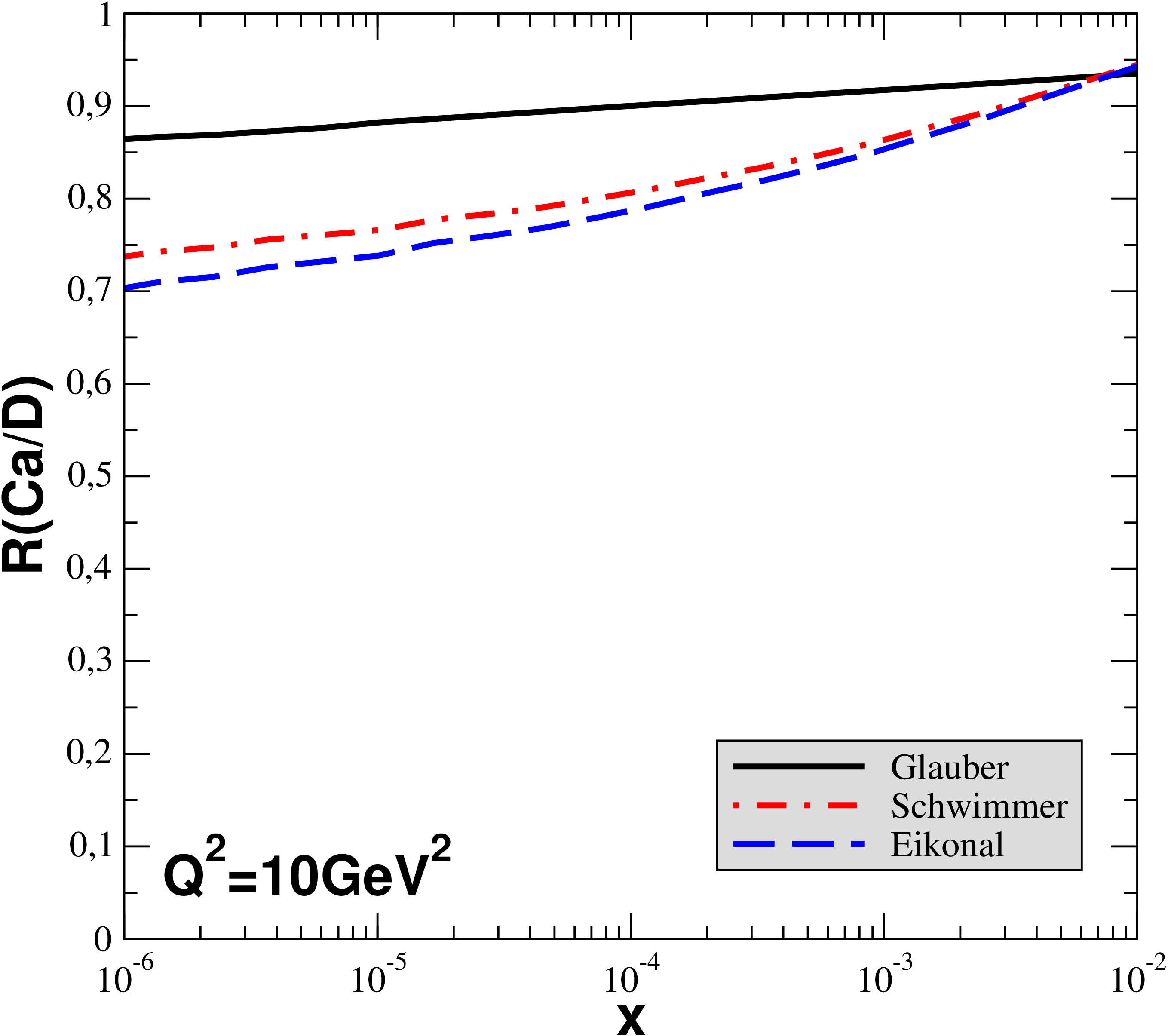}
\end{tabular}
\caption{Nuclear ratios $R_{Pb/D}$ (left panel) and $R_{Ca/D}$ (right panel) as a function of $x$ at $Q^2 = 10$ GeV$^2$. }
\label{fig3}
\end{figure}

In Fig. \ref{fig3} we present our predictions for the ratios $R_{Ca/D}$ and $R_{Pb/D}$ as a function of $x$ at  $Q^2 = $ 10 GeV$^2$. The behavior is similar to the one observed in the Fig. \ref{fig2}. The main point is that the differences between the predictions is not reduced significantly and this makes the 
discrimination between them  possible also at this value of $Q^2$.

A final comment is in order. The results shown in Figs. \ref{fig2} and \ref{fig3} demonstrate that there is a large 
uncertainty associated to the choice of unitarization scheme used to treat the multiple scatterings and that, in 
principle, an experimental analysis of the nuclear ratios can be useful to discriminate between these approaches. 

\begin{figure}
%\begin{tabular}{cc}
\vskip0.5cm
\includegraphics[scale=0.20]{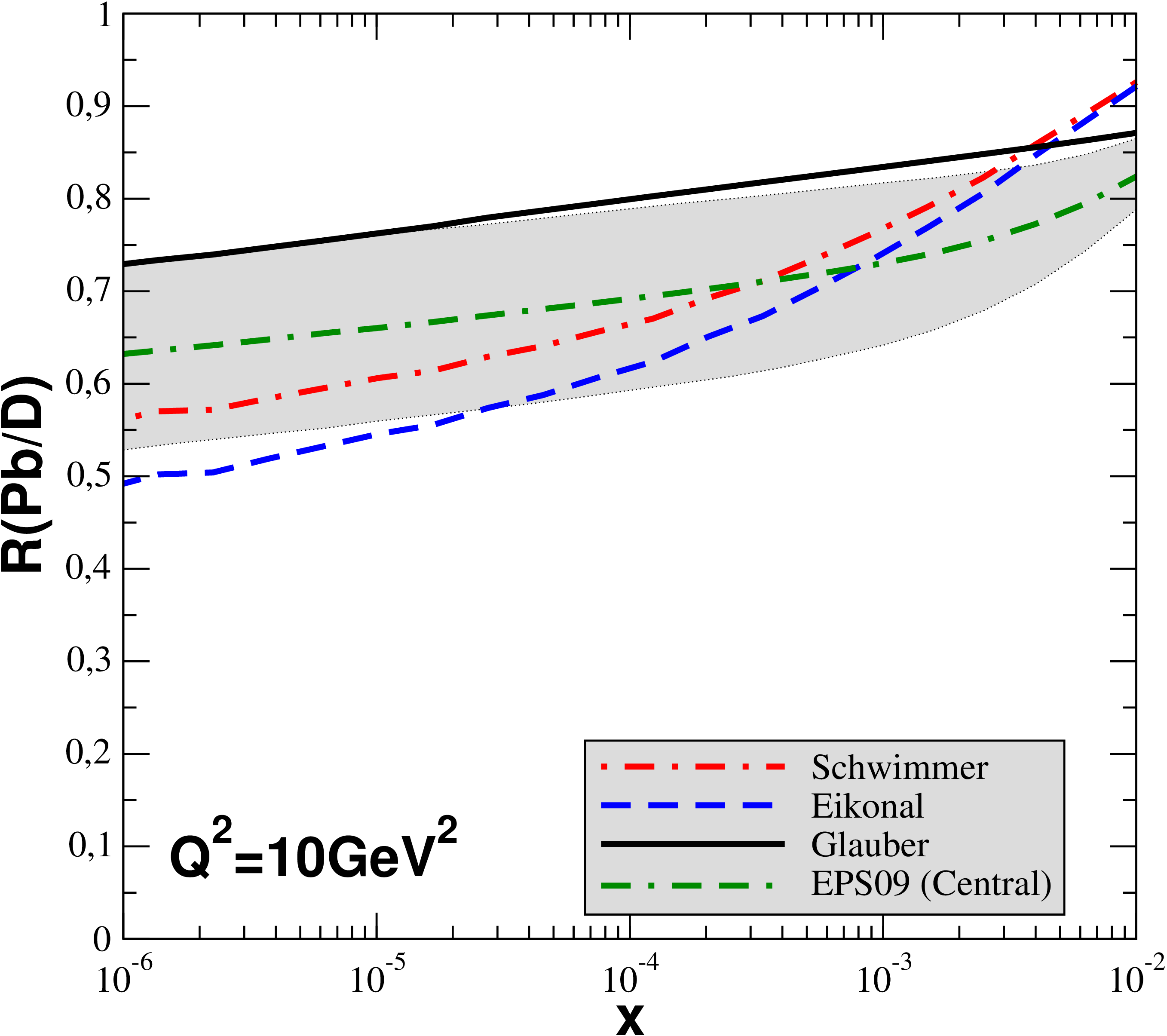}
%\psfig{file=hqpp_charm_rhic.eps,width=80mm} & \psfig{file=hqpp_bottom_rhic.eps,width=80mm}
%\end{tabular}
\caption{Predictions of the different models  discussed in the text. The dash-dash-dot line represents the 
central value of the prediction obtained with the EPS09 parametrization of the nuclear parton distribution functions.
The shaded band represents the theoretical error coming from the uncertainties in the EPS09 parametrization.}
\label{fig4}
\end{figure}

Another uncertainty present in the study of the nuclear effects is related to the transition between the 
linear and nonlinear regimes of the QCD dynamics. We do not know precisely in which  kinematical region  
the predictions obtained using the  linear DGLAP evolution cease to be valid. In Fig. \ref{fig4} we present a 
comparison of our predictions with those obtained using the EPS09 \cite{eps1} parametrization of  the nuclear 
parton distribution functions, which is based on a global fit of the current nuclear data using the DGLAP dynamics. 
As it can be seen, due to the large theoretical uncertainty  in the DGLAP prediction in the small-$x$ region, 
represented by the shaded band in the figure, it is not possible to draw any firm conclusion about which is the 
correct framework to describe this observable in future $eA$ colliders. This same conclusion was already obtained 
in \cite{erike_inclusive} in a  somewhat different approach. Consequently, the study of other observables, 
such as the nuclear diffractive structure function \cite{simone_eA,raju_eA} and nuclear vector meson production 
\cite{erike_exclusive,vector_eA}, should  also be considered  in  order to discriminate between 
the linear and nonlinear regimes. To summarize: in order to learn more about the 
unitarization schemes using the nuclear ratios  we must disentangle
the nonlinear and linear regimes of the QCD dynamics. Our estimates show that due to the large freedom present in the DGLAP analysis they predict similar 
magnitudes for the nuclear ratios, which implies that a combined analysis of several observables is necessary.

%-------------------------------------------------------------------------------
\section{Conclusion}
\label{conc}

The behaviour of the nuclear wave function at high energies provides fundamental information for the determination of the initial conditions in heavy ion collisions and particle production in collisions involving nuclei.
One of the main uncertainties is associated to the magnitude of the nuclear shadowing, which  comes mainly from  the  
way in which the multiple scattering  problem is treated and from  the modelling of the projectile - nucleon interaction.
Since a future EIC will probe  the shadowing region while keeping sufficiently large $Q^2$, new studies which determine the main sources of uncertainties in the predictions are necessary. In this work we compare three  frequently used approaches to estimate the nuclear shadowing in nuclear DIS.
As in these approaches  the nuclear cross section is completely determined once the interaction of the projectile with the nucleon is specified, we considered 
a single model (rcBK) as input of our calculations in order to quantify the theoretical uncertainty which comes from the choice of the unitarization model. In particular, we calculate the nuclear ratio between structure functions considering the Glauber, Schwimmer and Eikonal approaches down to very low-$x$ utilizing the rcBK results both for inclusive and diffractive cross sections in $\gamma^* p$ scattering.
Our results demonstrate that the current  experimental data at small $x$ are described successfully by the three approaches. However, the difference between their predictions becomes large in the kinematical region which will be probed in the future electron - ion colliders.

%-------------------------------------------------------------------------------
\section{Acknowledgments}

This work was partially financed by the Brazilian funding agencies CAPES, CNPq and FAPESP.

%-------------------------------------------------------------------------------

%--------------------------------------------------------------------------------

\end{document}